%% file: ms.tex
\documentclass[12pt,preprint]{aastex}
\newcommand{\EGRET}{{\it EGRET}\ }
\newcommand{\ROSAT}{{\it ROSAT}\ }
\newcommand{\GLAST}{{\it GLAST}\ }

\newenvironment{inlinefigure}{
\smallskip
\def\@captype{figure}
\noindent\begin{minipage}{0.999\linewidth}\begin{center}}
{\end{center}\end{minipage}\smallskip}

\slugcomment{}
\shorttitle{}
\shortauthors{Sowards-Emmerd et al.}

\begin{document}

\title{A Northern Survey of Gamma-Ray Blazar Candidates}
\author{David Sowards-Emmerd\altaffilmark{1}, Roger W. Romani, 
Peter F. Michelson\altaffilmark{1}, Stephen E. Healey\altaffilmark{1} \& 
Patrick L. Nolan}
\affil{Department of Physics, Stanford University, Stanford, CA 94305}
\altaffiltext{1}{also, Stanford Linear Accelerator Center, 
Stanford, CA 94039-4349}
\email{dse@darkmatter.stanford.edu, rwr@astro.stanford.edu, 
peterm@stanford.edu, sehealey@stanford.edu, pln@razzle.stanford.edu}

\begin{abstract}


In preparation for {\it GLAST}, we have compiled a sample of 
blazar candidates to increase the pool of well studied 
AGN from which \GLAST counterparts will be drawn.  
Sources were selected with our Figure of Merit (FoM) 
ranking; thus, they have radio and X-ray properties very 
similar to the \EGRET blazars.  Spectroscopic 
confirmation of these candidates is in progress, and more 
than 70\% of these objects have been identified 
as flat spectrum radio quasars and BL Lac objects.  We 
present $\sim$250 new optical blazar identifications based on McDonald 
Observatory spectroscopy, 224 with redshifts.  Of these, 167 are 
in our FoM-selected set.    

To motivate the $\gamma$-ray nature of these objects, 
we analyzed the current release of the \EGRET data for possible 
point sources at 
their radio positions.  We develop two distinct methods to     
combine multiple \EGRET observations of a sky position into a 
single detection significance.  We report a detection of 
the signal of the set of blazar candidates 
in the \EGRET data at the $>$ 3$\sigma$ level by 
both techniques.  We predict that the majority of these 
blazar candidates will be found by \GLAST due to its 
increased sensitivity, duty cycle and resolving power. 
\end{abstract}

\section{Introduction}

The \EGRET telescope on the Compton Gamma Ray Observatory 
({\it CGRO}) satellite detected 271 sources in a survey of 
the $\gamma$-ray ($\sim$100 MeV to 10 GeV) sky.  Counterparts for 
the majority of these sources have remained elusive, and until 
recently, roughly one quarter of these had been identified as 
blazars \citep{har99,mat01}.  By adopting a new 
technique, we have now pushed the identified fraction to 
$\sim$70\% above decl. $> -40^\circ, \vert b \vert > 10^\circ$, 
excluding the Galactic bulge (Sowards-Emmerd et al. 2003, 
hereafter SRM03; Sowards-Emmerd et al. 2004).  
The Large Area Telescope (LAT) on the Gamma-ray Large Area 
Space Telescope ({\it GLAST}) satellite, scheduled for launch in 
early 2007, is predicted to detect over an order of magnitude 
more blazars than its predecessor.  The improved 
sensitivity and resolving power of the LAT will provide 
more accurate source positions; however, the fainter sources will 
still only be localized to 5-10 arcminutes.  \GLAST will be 
instrumental in verifying the questionable \EGRET detections but will 
present an even larger set of its own marginal 
or unidentified $\gamma$-ray detections.  The current 
literature is lacking in well studied sources that would be 
realistic counterparts for the thousands of new \GLAST detections.

Blazars are jet-dominated active galaxies viewed 
close to the axis of a relativistic jet.  Blazar 
emission spans the entire range from radio to 
$\gamma$-rays and is often observed to be highly 
variable.  High-energy emission is thought to be the 
result of synchrotron self-Compton (SSC) upscattering \citep{up95};  
additionally, Compton upscattering of seed photons originating 
outside the jet is often invoked \citep{bot02}.

Under the current classification scheme, 
blazars are divided into two classes: flat spectrum 
radio quasars (FSRQs) and BL Lacartae objects (BL Lacs) \citep{up95}.
According to the current zoology, BL Lacs are a 
less luminous, more local population, drawn from a 
parent distribution of elliptical galaxies.  FSRQs, 
drawn from a quasar parent distribution, are generally 
higher power sources and are visible to much 
higher redshift.  Both classes typically exhibit compact 
radio cores with flat radio spectra ($\alpha < -0.5$, where 
$S_\nu\propto\nu^{-\alpha}$).  

To aid in the counterpart identification of \GLAST detections, we 
have created a blazar survey that seeks to identify a number 
of flat spectrum radio sources comparable to those expected from 
extrapolation of the 3EG \citep{har99} detections.  We select targets 
by applying our Figure of Merit (FoM) analysis, described 
in detail in SRM03, to objects in the Cosmic Lens All-Sky Survey 
(CLASS).  Excluding known 3EG counterparts, we have 
compiled a source list of 710 $\gamma$-ray blazar 
candidates.  The radio and X-ray properties of these 
sources are very similar to those of the 3EG blazars, and 
many are as bright or brighter in the radio.  As blazars 
are highly variable in $\gamma$-rays, we expect that many of 
these were simply inactive (in a `low state') when observed by 
{\it EGRET}.  We have found archival identifications and 
redshifts for nearly half of these sources in the current 
literature, and the remainder are being observed at 
McDonald Observatory.  We have contributed 241 new 
spectroscopic identifications in this paper.  This means 
that this northern sample is currently 74\% identified.

To further motivate the blazar nature of the survey 
sources, we have searched the current release of the \EGRET data 
for excess flux at the radio source positions.  To this end, 
we have developed two 
methods to rank $\gamma$-ray detections statistically
in the \EGRET data.  As a reality check on our technique, 
we calculate the equivalent significance of a subset of the 3EG 
sources.  

The selection criteria of the blazar candidate survey are outlined 
in Section 2.  The methods used to convert the \EGRET Test 
Statistic (TS) into a probability for an individual observation 
and to combine these observations into an overall significance of 
a $\gamma$-ray detection are described in Section 3.  In Section 
4, these two methods are applied to 3EG sources, survey sources, 
and the background sky.  A discussion of the merits of the 
two methods also appears there.  Section 5 details the 
spectroscopic observations of survey blazars.  Finally, Section 6 
presents a rudimentary comparison between the 3EG blazars and the 
blazar candidates presented in this survey. 

\section{Survey Selection}

\subsection{Figure of Merit Selection}

By comparing the properties of radio sources located within 3EG error 
ellipses (95\% confidence contours) to the background distribution, we 
have developed a new method of ranking counterparts of \EGRET $\gamma$-ray 
sources.  We believe that previous analyses relied too heavily 
on single dish 6 cm radio fluxes.  The recent release of the 8.4 GHz Cosmic 
Lens All-Sky Survey (CLASS) \citep{mye03} in the north provided uniform 
coverage over much of the northern sky.  In 
addition to more accurate core fluxes, non-simultaneous one-to-one 
spectral indices can also be extracted from the CLASS by pairing it with the 
1.4 GHz NRAO VLA Sky Survey (NVSS) \citep{con98}.  This 
mitigates the confusion in matching between the older single dish radio 
surveys and improves the spectral index determination.  Fluxes for CLASS 
sources were assigned by selecting the brightest source in each pointing 
and combining all the flux within a 1'' radius to include any compact jet 
features.  These CLASS sources 
were matched to the NVSS positions with a 4'' match radius.  Based on the 
number of sources in our survey, we expect on the order of one spurious 
match.  Finally, the 
\ROSAT All-Sky Survey (RASS) catalogs were also integrated into this data set 
to provide additional leverage in finding blazars.  

Examining these surveys in the context of 3EG, we quantified the 
correlation between flat spectrum radio sources and 3EG contours 
(SRM03), which took the form of a fractional excess, $n$, of 
sources (in a given flux or spectral index bin) found within the 
3EG positional 95\% confidence contours relative to the random 
background sources:
\begin{equation}
n = \frac{N_{\rm 3EG} - N_{\rm Random}}{N_{\rm 3EG}}
\end{equation}
\begin{equation}
{\rm FoM} = n_{\rm 8.4 \, GHz} \times n_{\alpha} \times n_{\rm X-ray} \times {\rm Positional \,\, Dependence} 
\end{equation}
where these functions are defined as:
\begin{equation}
%n_{\rm 8.4 \, GHz} = -3.47 + 2.45 \log \rm S_{8.4} - 0.338 (\log (\rm S_{8.4}))^2 
n_{\rm 8.4 \, GHz} = -3.47+2.45 {\rm Log}_{10}(s_{8.4})-0.34 [{\rm Log}_{10}(s_{8.4})]^2
\end{equation}
\begin{equation}
%n_{\alpha} = 0.187 - (0.350 \times \alpha) 
n_\alpha = {\rm Median}[0, 0.19-0.35\alpha_{8.4/1.4}, 0.4]
\end{equation}
\begin{equation}
%n_{\rm X-ray} = 0.987 + 0.411 \log \rm (Cnts) 
n_{\rm X-ray} = 0.5+ {\rm Median}[0, 0.49+0.41{\rm Log}_{10}(cps), 0.5]
\end{equation}
%where $n_{\rm 8.4 \, GHz}$ is truncated at 0. and 1., $n_{\alpha}$ is truncated 
%at 0. and 0.4, and $n_{\rm X-ray}$ is truncated at 0.5 and 1.0.  
The full analysis is described in detail in SRM03, including the positional 
dependence.  

Using the FoM method of selecting blazar candidates, we have created 
a survey tailored to find objects similar to the 3EG blazars.  
Radio candidates were selected from the 3.5 cm CLASS.  In this work, we 
exclude the positional dependence of the FoM and select objects based only 
on their intrinsic properties: radio flux, spectral index, and X-ray flux.  
We have selected sources such that their FoM would meet our original 
`plausible' ($<$18\% false positives) 3EG 
FoM limits if the source were located on a 3EG 95\% contour.  This leaves 
us with 710 sources selected from the CLASS survey that meet our criteria 
and were not previously associated with 3EG sources.  Properties of these 
sources are detailed in Table 1, the full version of which is available 
in the electronic version. The FoM 
selection effectively imposes a hard minumum on the CLASS radio flux 
of $\sim$140 mJy.

\subsection{X-ray Selected Sources}

Before the CLASS data were available, we selected target sources as 
an extension of the Deep X-ray Blazar Survey (DXRBS) \citep{per98} pushing 
fainter in radio flux.  Sources were selected from NVSS/GB6/RASS matches 
by the DXRBS criteria (keeping $\alpha<$0.7 and relaxing the flux cut).  
This means that a RASS X-ray detection 
was required for inclusion in the survey.  Application of the FoM 
technique shows that these sources are not strongly 
correlated with the 3EG positions.  Nevertheless, these sources are 
generally legitimate blazars, and so we include their spectroscopic 
properties for comparison.  
We note here that several of the 3EG blazars 
previously identified were not detected in the RASS, 
which in part led us to drop the X-ray detection requirement.  
Also, for high-power, low-peak blazars (FSRQs), the synchrotron peak 
is generally in the optical/IR so that the X-ray band, lying between 
the synchrotron and Compton components, can be quite faint.  In contrast, 
the higher peak BL Lacs often show bright synchrotron X-rays, but are less 
luminous in the GeV range.  
Most of the X-ray selected sample does not make the 
FoM cut since the radio fluxes are typically much fainter ($\sim$50 mJy) 
than the FoM-selected sources.  
For completeness, we tabulate redshifts and identifications in Table 2 
for the X-ray selected sources that did not meet the FoM criterion.  
Comparisons between properties of sources selected by this method 
and by the FoM method are presented in Section 6.

\section{Analysis of the $\gamma$-ray Data}

To further motivate the $\gamma$-ray nature of our blazar candidates, 
we investigate the possibility that 
these blazars produce flux detected by \EGRET but below the 3EG catalog 
threshold.  This is done through a reanalysis of the 
\EGRET data products, cleaned of the 3EG sources.  
We begin with the counts maps (photons), exposure maps, and maps 
modeling Galactic foreground emissions from the current release 
of the \EGRET data products.  These maps exist with a wide range of photon 
energy cuts. For this analysis we follow the standard procedure and 
only employ the {\it E} $>$ 100 MeV data.  
These data files are divided in time into individual `Viewing Periods' 
(VPs), which are pointings typically spanning a few days to a few 
weeks, and into `Cycles,' which comprise roughly 
a year's worth of observations.  Maps were also 
produced for all possible sequential combinations of these Cycles.

LIKE (v5.61 at the time of analysis) is the standard source 
detection routine and was developed specifically for \EGRET 
data \citep{mat96}.  LIKE computes a `Test Statistic' (TS) value 
through a maximum likelihood technique that compares the likelihood 
of detecting a source to the null likelihood (no source) for a 
given position through fits to the aforementioned data files.  Source 
positions, detection significances and fluxes can be extracted from 
the likelihood maps created by LIKE by searching for peaks in the maps.  
Once sources are detected, they can be modeled in the background or 
`cleaned,' and these `cleaned' maps can be searched again for the peaks in 
likelihood.  This procedure can be iterated down to a fixed TS threshold. 

At present, the 3EG catalog represents the most thorough 
analysis of the \EGRET data collected during survey mode
(Cycles P1-P4).  
Sources compiled in the 3EG catalog were selected by searching 
combinations of the \EGRET data for 
any $\sqrt{\rm TS}\geq 4$ detections ($\sqrt{\rm TS}\geq 5$ required 
for $\vert b\vert 
<10^\circ$).  The blocks of data searched for excess $\gamma$-rays using LIKE 
included VPs, a handful of combinations of sequential VPs, Cycles P1-P4, 
and the combinations of Cycles P12, P34 and P1234.  For a detailed 
description of the specifics of the 3EG catalog, 
we refer the reader to Hartman et al. (1999).  However, this 3EG detection 
threshold, required for inclusion 
in the catalog, does not account for the number of trials examined 
-- generally on the order of 15-35 -- nor does it account for the 
increase in significance 
associated with choosing the source position at the local TS 
maximum instead of using the more accurate radio counterpart position.  
Additionally, only sequential VPs 
were coadded and analyzed, although a randomly flaring source might be expected to appear in disjoint VPs.

\subsection{Searching the \EGRET Data}

From the previous analyses of the \EGRET data, we believe the $\gamma$-ray 
behavior of these blazars is often highly variable on timescales 
down to at least the length of \EGRET VPs.  Therefore, we select 
the individual VPs as the building blocks for our 
analysis.  However, we wish to avoid making 
any unnecessary assumptions about the $\gamma$-ray properties of these 
sources; that is, we do not want to be limited  to searching 
for sources that exhibit a single flare or, conversely, are steady 
in the $\gamma$-rays.  

In our search for correlation between this survey and $\gamma$-rays 
detected by {\it EGRET}, we chose to examine the significance of 
$\gamma$-ray detections at the precise radio positions of sources.
Using radio positions avoids `peaking up' the source positions on 
background fluctuations.  Obviously, this method generally returns a lower 
significance detection compared to using the position where the 
TS is a maximum ($\gamma$-ray position).  Therefore, 
our detection significance represents a conservative lower limit 
when compared with the 3EG rankings.  We believe this is the 
proper method for computing the flux or significance of a detection for 
\EGRET data when a precise position is known for a candidate counterpart. 

\subsection{Method}

We begin by analyzing the individual blocks of data: VPs ($\sim$few days), 
Cycles ($\sim$year), and combinations of Cycles.  Details of the beginning 
and ending dates and field centers of the VPs are summarized in Hartman 
et al. (1999).   
For each VP overlapping our region of interest, 
we create likelihood maps from the {\it E} $>$ 100 MeV maps using 
LIKE.  All 3EG sources in the field of view of a 
given viewing period are modeled (fluxes fit, not fixed).  Only 
data within $30^\circ$ ($19^\circ$ for the four `narrow-mode' 
pointings) of the field center are included in the 
analysis.  TS values representing the likelihood 
of source detection at a given point can be read directly off 
these $0.5^\circ \times 0.5^\circ$ gridded maps.

We analyze only the region covered by CLASS 
($0^\circ <$ decl. $< 75^\circ$ and 
$\vert b \vert > 10^\circ$).  Additionally, we remove the 
Galactic bulge to a radius of $30^\circ$ from the Galactic center 
due to the high density of 3EG 
detections in the bulge.  Finally, regions of $2^\circ$ radius around 
3EG sources were excluded from the analysis.  LIKE tends to oversubtract 
sources when they are modeled, leading to an excess of low TS
values relative to a random background;  that is, when a point source 
is modeled by LIKE, the resulting likelihood distribution in that 
region is 
not consistent with random background and is instead almost entirely 
filled with 
TS $<$ 1 values.   If sources were properly cleaned, one would expect 
to find a random distribution (e.g., approximately one TS $\geq$ 4 
($2\sigma$) point for every $\sim$40 
points) where a source was modeled.  This oversubtraction problem 
becomes insignificant at approximately 2$^\circ$ from \EGRET sources.  
A precise treatment of this issue would involve making cuts 
around each 3EG source based on its flux in each individual 
viewing period; however, we believe that a simple 2$^\circ$ cut is 
satisfactory for the purpose of this work.  The subset remaining 
after applying these cuts contains 578 of the original 710 objects.

\subsection{Converting TS into a Probability}

With TS values in hand, we require a means to transform a set of 
TS into a meaningful probability of a $\gamma$-ray excess at a given 
position.  We consider three approaches to 
converting TS to a properly normalized 
probability.  The trivial method employed in Mattox et al. (1996) is 
simply to assume that TS is distributed like $\chi^2_1/2$.  From this,  
one can simply convert via $\sqrt(\rm TS) \rightarrow \sigma$.  
This predicts that 
exactly 50\% of the data points in a blank field should have TS = 0.  
However, in the 
northern sky data cleaned of the 3EG sources, even excluding 
2$^\circ$ around the positions of the 3EG sources, 
we find $\sim$59\% of the points have TS = 0.  We believe that this 
is largely due to overfitting of \EGRET point sources and the free 
normalization of the diffuse background flux.  This therefore 
produces a large discrepancy in any analysis that includes 
low TS values.  This method may be reasonable when applied 
to large TS values; however, in searching for faint detections 
it is clearly inadequate.  

The following two methods treat the cleaned \EGRET data as the starting 
point and use the observed distribution of TS values to determine the 
corresponding probability.  This is done by summing up the number of 
positions in a given VP with TS greater than a given value and 
normalizing this to the total number of points in the viewing period:
Prob($>$TS) = \#($>$TS)/\#(Total).  
This probability function is only applicable to points in a single VP, 
so this procedure is repeated for each VP and Cycle.  
The overall number of positions in a VP is the 
limiting factor here, and due to the spatial cuts listed above, this 
number can be as small as a few hundred.  
For small TS values, this method, while heuristic, is the most robust.

To extend the analysis to lower probabilities (larger TS), we can 
concatenate the VP TS distributions and compute the probability 
distribution for all VPs at once.  In doing this, we are able to increase the 
number of points considered by two orders of magnitude, and 
hence the resulting range of probabilities can be extended by a 
comparable amount.  This approach averages out the VP-to-VP 
variations in the low-TS end of the distributions.  
This is checked by comparing with the results of the individual VP 
method; little 
difference is observed between the resulting final probabilities 
for low-significance sources.  Thus, as one would expect, the background 
fluctuations average out in the sets of overlapping VPs.  
Since these differences do average out, 
we chose to adopt this method because it allows the larger TS values to 
be treated uniformly and it extends the probability range by two orders of 
magnitude.  By selecting this 
occurrence-based method over $\chi^2$, we are treating 
the data as though we are starting from scratch with the 
3EG-cleaned maps.  Examples of these distributions are shown 
in Figure 1. 

\clearpage
\begin{inlinefigure}
\figurenum{1}
\scalebox{1.0}{\rotatebox{0}{\plotone{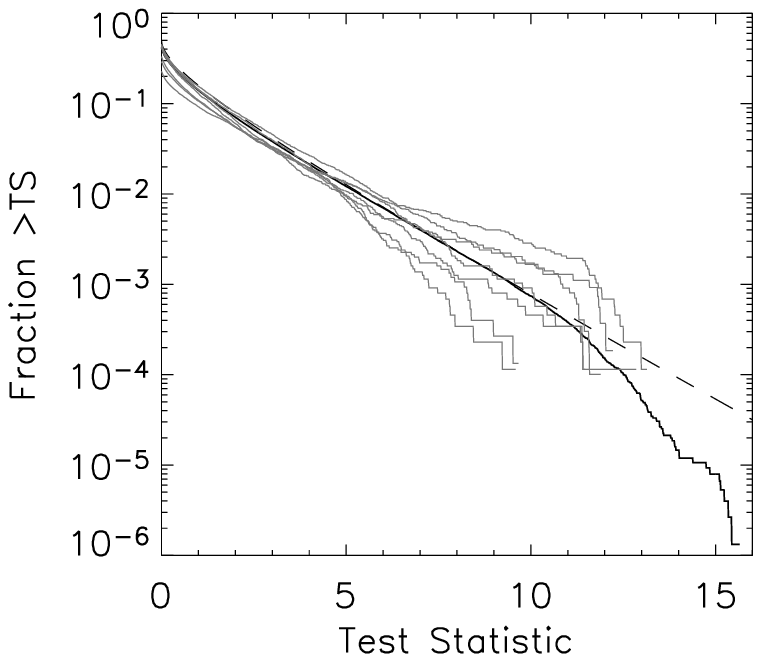}}}
\figcaption{The distribution of TS for 
seven individual VPs (gray) and the concatenation of 
all 169 VPs (solid black).  This may be compared with the 
$\chi^2/2$ distribution (dashed black).  Note the large discrepancy 
between these distributions as TS approaches zero.  This effect 
is averaged out in the concatenated distribution.  Additionally, 
note the large variation in the maximum TS values found in 
individual VPs.}
\label{Fig1}
\end{inlinefigure}
\clearpage

\subsection{Combining the Information}

For each test position, we now have a set of probabilities 
corresponding to the individual VP observations.  Our task now is 
to integrate these probability estimates into a single detection significance.  
We investigate two methods.  The first method takes an agnostic 
stance on the temporal behavior of a source and 
combines all the information available on the VP timescale.  
We believe that it should be possible to extract almost 
all of the information available in the Cycle and combination maps by 
combining information from the individual VPs.  
We create the `combined' probability at a given position by multiplying the 
individual VP probabilities in which that position was observed.  Clearly, 
the resulting probability is strongly dependent on the number of VPs, 
so we must renormalize.  We do so by integrating over the 
volume of {\it N}-dimensional combined probability space where the combined 
probability is less than or equal to that of the source.  That is, we 
determine how likely it is to build up the combined probability in question 
from {\it N} observations of random TS values.  
This problem can be solved analytically, and the final, 
normalized probability can be written as:
\begin{equation}
P_{\rm norm} = p \sum_{i=0}^{N-1} (-1)^i \frac{\ln^i p}{i!}
\end{equation}
where {\it p} is the product of VP probabilities at a given position.  
Using this procedure, we can compare positions observed in different 
numbers of viewing periods.  We designate this probability as the `Product' 
probability.  

In the second method, for each position, we create a set of TS 
values extracted from the likelihood maps from VPs, Cycles, 
and combinations of Cycles (P12, P34, P1234).  We include combinations 
of Cycles in this case because we are searching for the single 
strongest detection, be it steady or flaring, and discarding the 
remaining observations.  The 
normalized probability is then calculated by selecting the 
maximum TS (minimum probability) from this set, converting it 
into a probability, and then normalizing this probability 
for the number of trials using the equation:
\begin{equation}
P_{\rm peak}=1 - (1-p_{\rm min})^N
\end{equation}
where {\it N} is the number of cases examined, and $p_{\rm min}$ 
is the minimum probability in the set of observations.  This procedure, 
refered to as the `Peak' method, excels at 
finding sources that were detected strongly in a single 
viewing period.  This is the preferred method for finding 
blazars if the flaring timescale is comparable to a VP and 
if they flare only once during the set of observations.  This 
technique corresponds most closely to the standard 3EG analysis.

\subsubsection{Call for a 4EG Catalog}

Based on the \EGRET analysis carried out in this work, we believe 
that a fourth \EGRET catalog is essential to preparation 
for {\it GLAST}, due to reprocessing of the data since the 
creation of the 3EG catalog.  In analyzing the current release of the \EGRET 
data (assuming 3EG positions for the sources, but re-modeling the 
fluxes and Test Statistics), we have found 
several sources that would have been 
dropped from the list after reanalysis.  One particularly 
bad example is J1227+4302, whose peak detection drops down to 
$\sqrt{\rm TS}=1.9$.  
A dramatic decrease in photons in the vicinity of this source is 
visible in comparing the current and 3EG {\it E} $>$ 100 MeV raw 
count maps.  
When comparing the distribution of TS values extracted from the 
current \EGRET data products (cossc.gsfc.nasa.gov - 'Current 
Date' in the old and current maps are 27/8/96 and 27/3/01 
respectively) to the 3EG 
catalog, an overall drop in detection significance is apparent.  
While the overall exposure in P1234 has increased in the current 
reprocessing, the counts have decreased in both the $<$100 MeV 
and $>$100 MeV maps.  More quantitatively, the counts decreased 
by 2.3\% and 0.6\% in the $>$100 MeV and $<$100 MeV maps 
respectively, while the exposure increased by 0.4\% and 0.3\%.  
Additionally, improved Galactic foreground emission models may further 
alter the extracted source list. 

\section{Application of the Method}

\subsection{Low Significance 3EG Sources}

As a reality check, we compute the detection significance 
by both methods for a subset of the 3EG detections.
In order to apply these methods to the 3EG sources, 
we first have to generate TS maps from which 
these sources are not excluded.  We create two sets 
of these maps: one where 3EG sources with $\sigma>$ 4.75 
are `cleaned' (modeled and removed), and one where $\sigma>$ 4.25 
sources have been cleaned.  
From these maps, we simply read off the TS values for each VP/Cycle at the 
3EG positions of the $\sigma<$ 4.25 and $\sigma<$ 4.75 sources.  This 
differs from the traditional \EGRET method of modeling all sources 
simultaneously (which we have also done).  Note that here the source 
positions used were the 3EG $\gamma$-ray positions, not the positions 
of radio counterparts, which should increase the apparent significance.
The distributions of Peak and Product probabilities for this sample are 
plotted in Figure 2.

\clearpage
\begin{inlinefigure}
\figurenum{2}
\scalebox{1.0}{\rotatebox{0}{\plotone{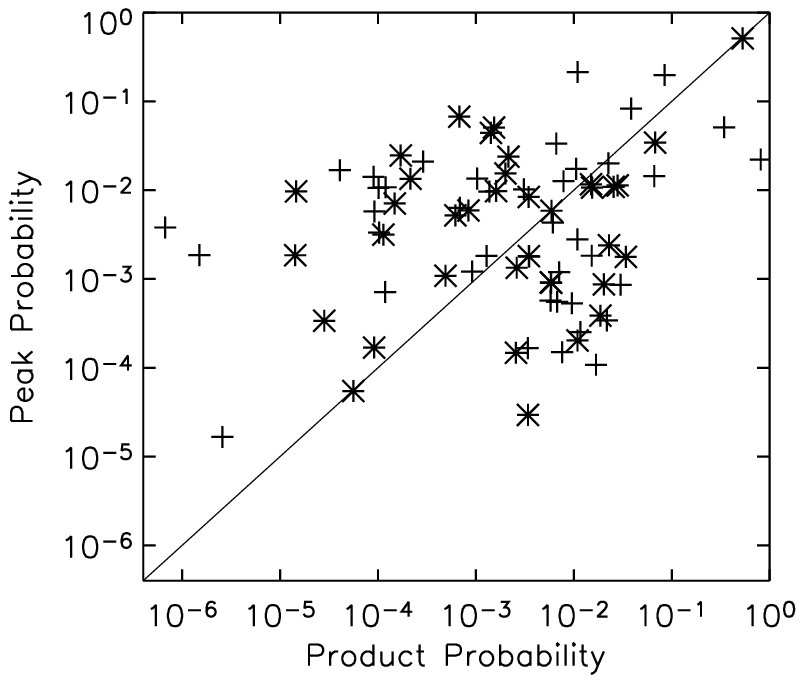}}}
\figcaption{The probabilities are plotted for faint 3EG 
sources for both the Product and Peak methods.  Pluses 
and stars represent $4.25<\sqrt{\rm TS}<4.75$ and $\sqrt{\rm TS}
<4.25$ sources respectively.  The threshold for inclusion in the 3EG 
catalog appears to be $\sim$2$\sigma$ after accounting for the 
number of trials.  
Both methods show a wide range of statistical significance, strongly 
affected by the number of trials.  For the selected sources, this 
dominates the range of the Peak method probability.  
The two methods exhibit large differences in $\gamma$-ray 
probabilities for individual sources and we believe these methods are 
complementary.  The lowest significance point is J1227+4302, a 
source whose peak TS dropped from $\sim$16 to $\sim$4 in the 
current release of the data.}
\label{Fig2}
\end{inlinefigure}
\clearpage

Since we have correctly normalized for the number of trials, we expect 
that the probabilities will be significantly lower than 
the 3EG.  The significance threshold of sources in the 3EG catalog is 
in fact closer to 2$\sigma$ when we account for the number of trials.  
As noted above, in addition to the decrease in 
significance due to the normalization process, we also find a 
systematic drop in TS values for these sources in the current release 
of the data, arguing that a fourth \EGRET catalog should be created.

\subsection{The 3EG-Cleaned Background Distribution}

In order to determine whether the Peak and Product methods produce 
reasonably normalized probabilities, we created a `background sample' 
in the form of a grid of $0.5^\circ \times 0.5^\circ$ pixels spanning 
$0^\circ<$decl.$<75^\circ$, $\vert b \vert > 10^\circ$ and a 
minimum of $2^\circ$ away from modeled 3EG sources.  We then filled in 
this grid with the actual TS values from the VP and Cycle maps.  Since we do 
not model the `$3\sigma$ 3EG' sources or anything fainter, we 
expect real sources to be present in this background.  However, 
we anticipate that the low-TS values should be 
distributed as true random background.  This is nearly the case 
for the Product background distribution; however, the Peak 
background distribution is far from random.  Therefore, 
to compare these methods on equal footing, we must renormalize 
these probabilities.

We choose to renormalize these probability 
distributions by renormalizing the background probability distribution 
to a uniform probability distribution.  
This is accomplished by computing the occurrence-based 
probability for survey sources from the background distribution.  
That is, for a given Peak or Product probability, we total the number 
of points in the respective background distribution with this 
probability or lower, divide this by the total number of points 
in the background, and assign this as our renormalized probability.  
Because we know that this background includes numerous sources, all of 
which span tens of pixels, this renormalization produces a very 
conservative measure of the likelihood of $\gamma$-ray detection.  
In effect, we are calculating the probability that a source is detected 
more strongly than a random sky position, where the sky contains 
unmodeled sources.

This procedure would not be necessary if there were no spatial 
correlations in the data.  This is seen through the clustering 
of TS = 0 detections seen at a significant fraction of positions 
on the sky.  Of the 61,866 positions ($0.5^\circ \times 0.5^\circ$ 
pixels) that we consider in the north, we find that 
more than 1000 positions have TS = 0 for every observation.  This is a 
very significant excess over the $\sim$dozen such points we would 
expect to find by chance.  We have confirmed via Monte Carlo that 
combining an arbitrary number of uniformly distributed probabilities by 
either the Peak or Product method results in a uniform distribution of 
probabilities, demonstrating that these correlations do exist in the 
data and are not intrinsic to the computation methods.

\subsection{Source Detection}

Now that the methods are well understood, we search the \EGRET data 
for the FoM-selected blazar candidates.  
Each technique has its own merits, and we do not deem either method 
superior.  Both the Peak and Product ranking methods 
produce a significant detection of the blazar candidates above the 
background.  The Peak method excels at finding sources that flare 
up just once, especially when they sputter around below background 
for the remainder of the observations.  By not including all 
possible combinations of data -- we only include individual VPs, P1, 
P2, P3, P4, P12, P34, and P1234 -- the Peak method misses sources 
that exhibit multiple flares spanning several cycles (e.g., flaring in 
P1 and P3).  Conversely, the Product 
method is more efficient at finding sources that are detected multiple 
times, especially sources where the largest TS is found in P1234.  
The Product method is much more sensitive to the number of VPs in 
which a position was observed.  

\clearpage
\begin{inlinefigure}
\figurenum{3}
\scalebox{1.0}{\rotatebox{0}{\plotone{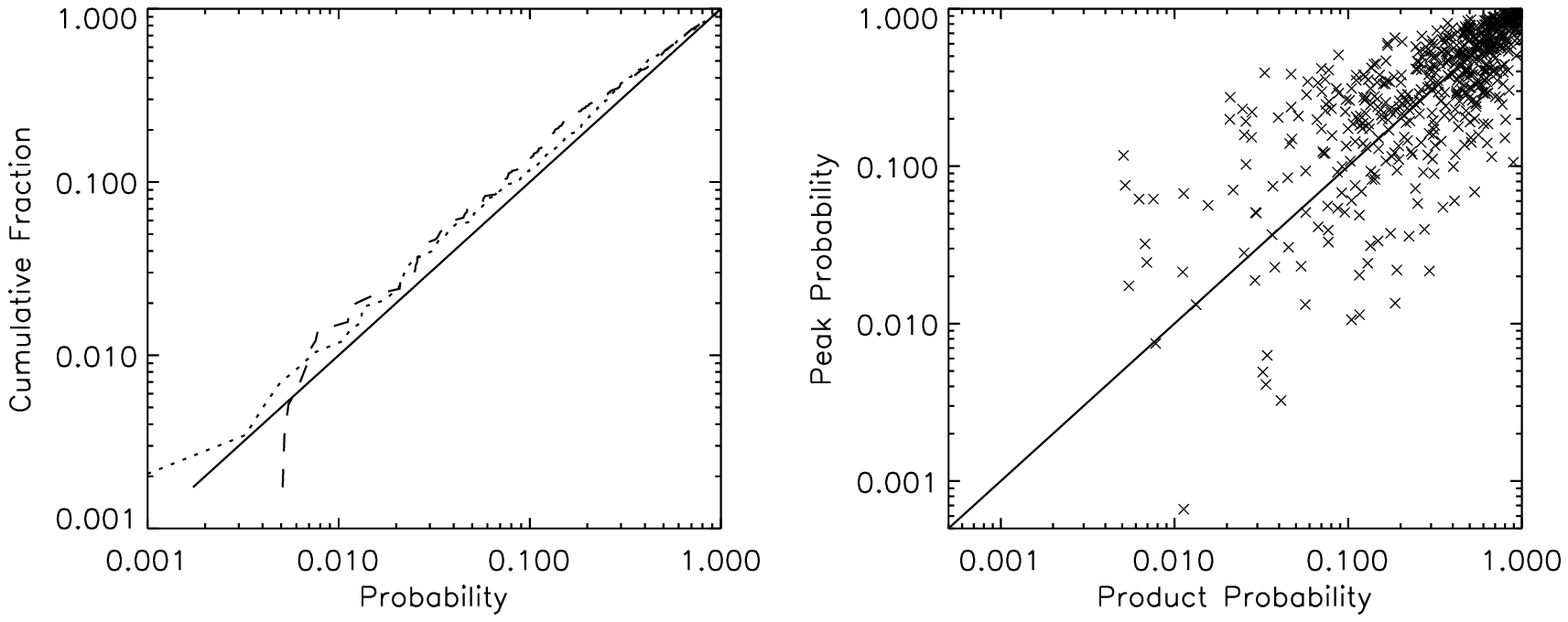}}}
\figcaption{Left: The cumulative distribution of probabilities for the 
Peak method (dotted) and Product method (dashed) are compared to 
the expected random distribution (solid).  Right: Scatter plot contrasting 
the Peak and Product methods of calculating the $\gamma$-ray likelihood.  
Sources detected with higher significance by the Product method appear above 
the solid line and those detected more strongly by the Peak method lie below 
this line.  Generally, most sources are best detected by the product method, 
however a few flaring sources show up best as Peak method 
detections at the bottom of the plot.} 
\label{Fig3}
\end{inlinefigure}
\clearpage

As can be seen in Figure 3, both probability distributions deviate 
significantly from a uniform random distribution.  We quantify this 
significance in two ways.  First, we compare the distribution of 
probabilities for the blazar candidates computed by each method to the 
background distribution via the K-S test.  The resulting probabilities 
are 0.00021 (3.7$\sigma$) and 0.0018 (3.1$\sigma$) for the Peak and 
Product methods respectively.

Second, by setting several detection thresholds, we compile the number 
of sources selected by each method.  These are shown in Table 3.  
In Table 1, a sample page of the complete table, which is available 
electronically, the Peak and Product probabilities are tabulated 
for all sources that survive the spatial cuts.

\subsection{Does $\gamma$-ray Probability Correlate With FoM?}

Since we believe that the FoM is selecting {\it EGRET}-like blazars, 
we expect a correlation between the FoM and the $\gamma$-ray 
probabilities at a given location.  
It is unclear whether blazars of comparable 
radio brightness to the 3EG blazars are not strongly detected because 
they were not actively flaring in $\gamma$-rays 
during \EGRET observations or because of a difference in the opening angle 
of the radio and $\gamma$-ray beams.  Either of these effects 
decreases the correlation between the FoM and the $\gamma$-ray 
detection probability.  Of course, since these survey sources are 
typically found to be very faint in $\gamma$-rays and background 
fluctuations are comparable in magnitude, we do not expect to find a 
strong correlation.   

We examine the FoM/$\gamma$-ray probability correlation in the data 
in three ways.  First, 
for each probability scheme, we split the survey data by FoM into 
two nearly equal parts (at FoM = 0.075) and compare the resulting 
distributions to the background.  In the Peak case, both sets 
show similar significance in departure from the background as 
measured by the K-S test: FoM $>0.075 \rightarrow 0.034$, FoM $<0.075 
\rightarrow 0.0095$.  
However, for the Product method, the larger FoM set is much more 
significant: FoM $>0.075 \rightarrow 0.0027$, FoM $<0.075 \rightarrow 
0.28$. 
 
Alternatively, we compute the Spearman rank correlation 
for both the Peak and Product methods against the FoM.  The resulting 
coefficients 
(significances) are -0.07 (1.7$\sigma$) and -0.02 (0.54$\sigma$) for 
the Product and 
Peak methods respectively.  The correlation has the correct sign for 
both methods since we expect the smaller probabilities for large FoMs.  
For the Product method, the correlation is mildly significant; however, 
for the Peak method, the results are insignificant.

Finally, we create a $\gamma$-ray selected subset of the survey by 
selecting sources detected at the 95\% (2$\sigma$) level by either the 
Peak or Product method and then examining the distributions of 
corresponding FoM values.  Applying this cut, we find 41 sources by 
the Product method (34 Peak), and the mean FoM value calculated for 
these sources is 0.098 (0.089) compared to the average value of 
0.083 for the whole survey.  Thus $\gamma$-ray selection appears 
to select brighter, flatter radio sources than average, and the 
Product method is more effective than the Peak method.  In all three 
of the above tests, the strongest correlation is found between the FoM 
and the Product method.

\section{Optical Follow-up}

Of our 710 survey blazar candidates, nearly half have archival 
classification and assigned redshifts.  Information for these objects 
was extracted primarily from the Eleventh Edition of the Quasar Catalog 
(Veron-Cetty et al. 2003) and the Early through 3rd data releases of the 
Sloan Digital 
Sky Survey \citep{aba04}.  A handful of objects also appear in 
the optically bright CLASS Blazar Survey \citep{cac02}, the only other 
blazar survey drawn from CLASS sources.  
As a follow-up, the remaining positions were 
queried in the SIMBAD database.  The remaining objects are the 
target of spectroscopic observations at McDonald Observatory. 

\subsection{Hobby-Eberly Telescope/LRS Spectroscopy}

Spectroscopic observations were made using the Marcario Low 
Resolution Spectrograph (LRS) 
(Hill et al. 1998) on the 9.2 m Hobby-Eberly Telescope (HET) 
(Ramsey et al. 1998).  These 
targets were observed in regular queue operations from March 2002 to 
October 2004.  Exposures ranged from 2 $\times$ 300 s to well over an hour 
spanning several visits.  Image quality generally fell in the range of 
1.5'' to 2.5''.  Observations employed a 300 lines mm$^{-1}$ 
grating and a 2'' slit for a dispersion of 4 \AA ~per (2$\times$2 
binned) pixel and an effective resolution of 16 \AA ~covering 
approximately $\lambda\lambda 4200-10000$. 
 
Standard IRAF CCD reductions, calibrations, and optimal extraction were 
performed.  Redshifts were derived whenever possible via cross-correlation 
analysis with AGN and galactic templates using the IRAF RVSAO package.  
Images of the extracted 1-D spectra are available online at 
\url{astro.stanford.edu/northernblazars.html}.

We do not yet have spectroscopic data for 186 of the blazar 
candidates and are still observing these objects with the HET.  

\subsection{2.7 m/IGI Spectroscopy}

A handful of objects were observed with the 2.7 m Harlan J. Smith telescope 
at McDonald Observatory.  Filler observations were made during an 
observing run from 25-29 July 2003, primarily targeting 
southern 3EG counterparts.  The Imaging Grism Instrument (IGI) 
spectrograph (Gary Hill) was used with a 6000 \AA ~Grism and a 50 mm lens. 
A 2'' slit was employed for the first 1.5 nights and a 2.5'' slit 
thereafter due to generally poor imaging.  The imaging fluctuated around 
1.5'' to 2'' on the best night, but typically held around 2'' to 2.5''.  
The IGI setup covered a smaller wavelength range than the HET/LRS, namely 
$\lambda\lambda 4250-8500$.  Due to a wide range of observing conditions 
and source brightnesses, total exposures ranged from 300 to 3600 seconds.

\subsection{Classification}

As of this publication, we have obtained 167 new spectroscopic 
identifications for objects selected by the FoM method.  The 
redshifts and basic properties of these new blazar identifications 
are listed in Table 1.  Sources observed with the HET or 2.7 m were 
classified based on their observed 
spectra, S/N permitting.  The observed sources fall into the following four 
categories:  BL Lac objects, passive elliptical galaxies, flat spectrum radio 
quasars, and narrow line radio galaxies.  The featureless spectra and 
spectra with strong absorption lines were segregated into two categories: 
BL Lac objects and passive elliptical galaxies.  BL Lac objects are 
defined here by the following properties (Marcha et al. 1996): the break 
contrast $ =(\frac{f^+ - f^-}{f^+}) < 0.4$, where $f^+$ and $f^-$ are the 
fluxes 
redward and blueward of the H/K break, and the rest-frame equivalent width 
of the strongest emission line detected must be less than 5 \AA.  The 
remainder of the absorption line objects 
are classified as passive elliptical galaxies, although these appear to 
have active nuclei.  For a handful of objects, only narrow emission lines 
(kinematic width $<$ 1000 km/s) were observed and these were classified as 
narrow line radio galaxies (NLRG).  All other emission 
line objects were classified as flat spectrum radio quasars.

Archival data were classified based on the source classes 
of the 11th 
Veron-Cetty/Veron Quasar Catalogue, Sloan Digital Sky Survey and 
SIMBAD queries.  
Objects listed in the Veron catalog as Type 1-1.5 Seyfert galaxies, 
active galactic nuclei and quasars were 
classified as flat spectrum radio quasars. Seyfert Type 1.9 and above 
were classified as narrow line radio galaxies. 

\subsection{Survey Efficiency}

We find that the FoM method is extremely efficient at targeting blazars.
In the subset of the sample with reliable identifications, over 95\% of 
sources are identified as FSRQs or BL Lacs.  Thus we find a factor of 
3 fewer spurious sources by this method than in our 
X-ray selected pilot survey.  Interestingly, $\sim$25\% of the 
non-blazars found were X-ray detected, suggesting that requiring X-ray 
detection for an AGN survey is more likely to select NLRGs and passive 
ellipticals.

\subsection{Interesting Spectra}

This survey has produced several interesting objects. 
The most noteworthy is J0906+6930, which we believe to be the highest 
redshift blazar to date at {\it z} = 5.47 \citep{rom04}.  This is 
also the highest redshift radio-selected object to date 
and the brightest object at this redshift at 8.4 GHz.  
Preliminary 2 cm VLBA snapshots find a jet-like 
structure, providing more evidence for the blazar nature of this 
object, and compact flux was also detected at 7 mm at the 40 mJy level.  

One FSRQ, J1618+0819, exhibits extremely broad lines (FWHM$(v_{kin})>$ 17,000 
km/s), where the Balmer lines are clearly multi-peaked, on top of a 
very strong continuum.  J1855+3742, another FSRQ, shows a strong CII 
line (2326 \AA) not generally observed in blazars, the 
equivalent width of which is comparable to that of MgII.  
Additionally, we find several new bright featureless BL Lac objects.  
Spectra of these objects are shown in Figure 4.

\clearpage
\begin{inlinefigure}
\figurenum{4}
\scalebox{1.0}{\rotatebox{0}{\plotone{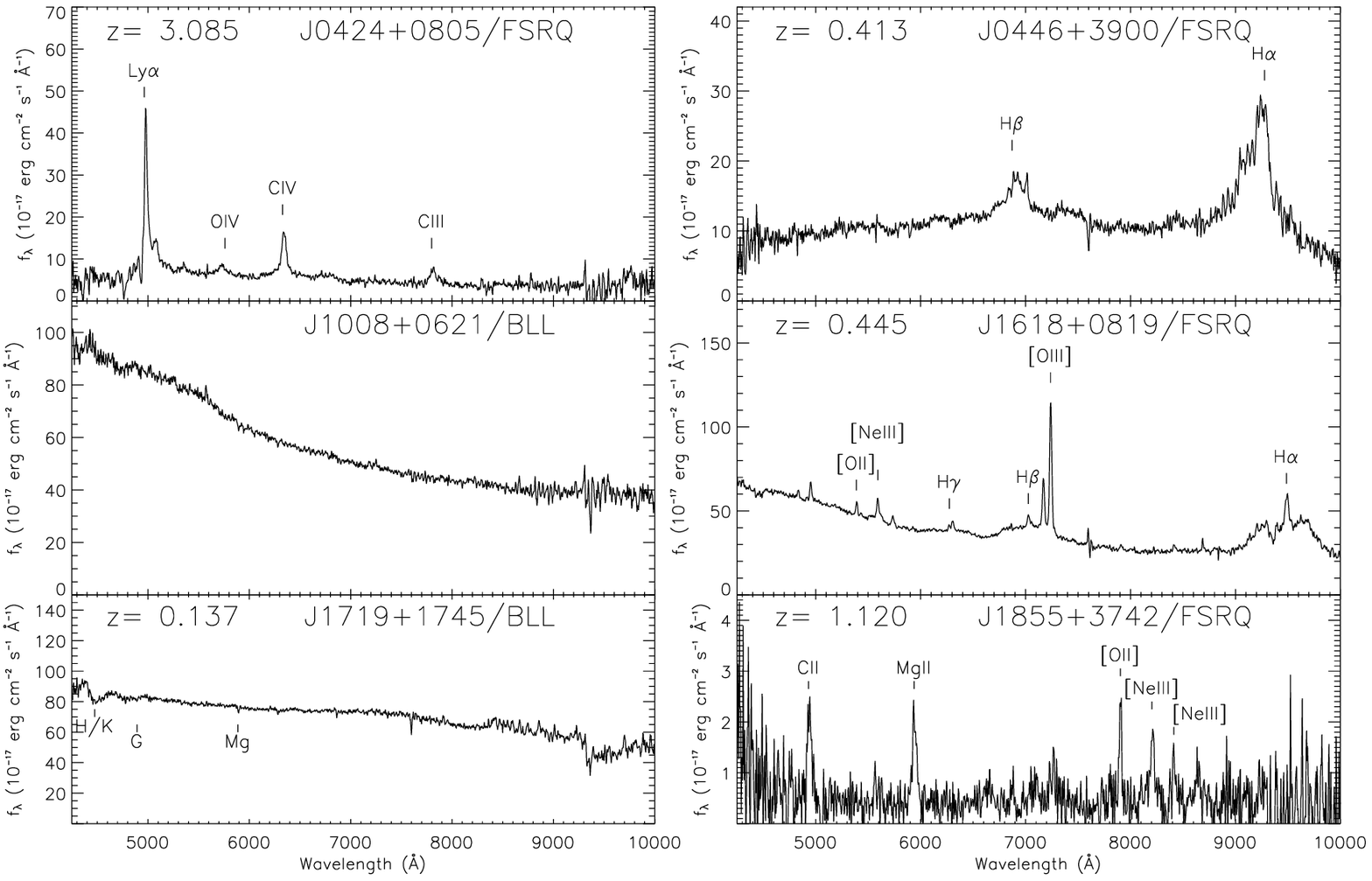}}}
\figcaption{Sample Spectra:
J0424+0805 -- A moderately high redshift FSRQ, 
J1008+0621 -- A typical featureless BL Lac object, 
J1719+1745 -- A source classified as BL Lac in the literature without 
redshift - we have determined a robust redshift from our new 
spectroscopy and confirm the BL Lac ID, 
J0446+3900 and J1618+0819 -- FSRQs with extremely broad Balmer lines 
(multiple components), and
J1855+3742 -- A weak-continuum FSRQ; note the unusual occurance of the 
CII emission line.} 
\label{Fig4}
\end{inlinefigure}
\clearpage

\section{Source Populations}

We can now compare this FoM-selected sample with the 3EG sample.  We believe 
that this survey is composed of both a fainter extension of the 3EG as well 
as sources that would have been detected by \EGRET had they not been in a 
low-flux state when observed.  We examine the redshift distribution and 
source population of these samples as well as the effects of X-ray 
selection.

\subsection{Redshift Distribution}

The primary difference between sources in this sample and the 3EG 
counterparts is simply the proximity to strong \EGRET $\gamma$-ray 
detections.  The redshift distributions of these two samples are 
indistinguishable by the K-S test (Prob$_{\rm K-S}=0.19$).  For 
comparison, the redshift histogram of FSRQs from this work 
is plotted against the distribution of 3EG counterparts (SRM03, SRM04) 
in Figure 5.  
Only the FSRQs are considered in this exercise since for the BL Lacs we 
have a much smaller number of redshifts, and the redshift measurements are 
much less uniform than in the well observed 3EG sample.  
The X-ray detected survey sources are on average less distant 
than the non-X-ray detected sources.  The mean redshifts of the FoM 
survey blazar candidates are 1.08 and 1.52 for the X-ray detected and the 
non-X-ray detected sources respectively, averaging to 1.37 overall.  
Our DXRBS-like survey, which required X-ray detections, had an even lower mean 
redshift of 0.84.  The 3EG blazar 
counterparts fall at a slightly lower redshift of {\it z} = 1.32 on average but 
are consistent with the FoM-selected distribution.  
Not surprisingly, requiring X-ray detections selects a more local 
population of sources since the RASS is a relatively shallow survey. 
The set of blazar candidates detected in the RASS were on average 50\% 
brighter at 8.4 GHz than those not seen in X-rays.  Additionally, the 
X-ray detected FoM-selected blazar candidates are 1.3 magnitudes (USNO 
B1.0 R2 magnitude)(Monet et al. 2003) brighter optically than the rest.  
These tend to be 
`high peak' blazars (HBL) with the synchrotron peak extending to the 
optical and X-ray \cite{up95}.  These will be relatively low luminosity 
$\gamma$-ray sources.

\clearpage
\begin{inlinefigure}
\figurenum{5}
\scalebox{1.0}{\rotatebox{0}{\plotone{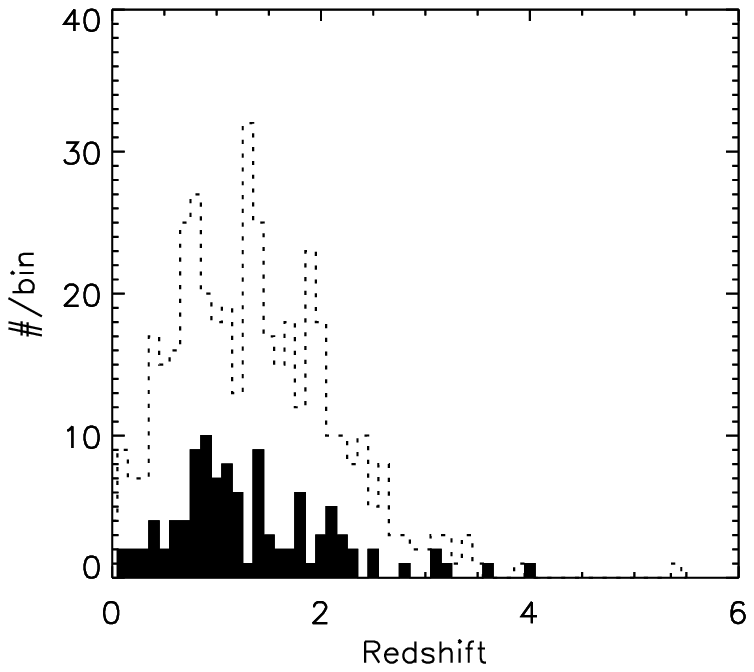}}}
\figcaption{Redshift histograms of the 3EG FSRQs (filled) 
and the survey FSRQs (dotted line).  Comparing the two 
distributions via the K-S test results in a probability of 
0.19 that they are drawn from the same parent distribution.}
\label{Fig5}
\end{inlinefigure}
\clearpage

\subsection{BL Lac Fraction}

In the 3EG sample, above decl. = -40$^\circ$ and away from the Galaxy, we found 
$\sim$18\% BL Lac objects and $\sim$78\% flat spectrum radio quasars.  
To date, in the FoM-selected survey we find a lower BL Lac fraction 
of $\sim$13.6\%.  This likely stems from the fact that the 3EG 
sample is nearly complete, while the FoM sample only has identifications 
for $\sim$74\% of the sources, where the majority of these IDs are archival.  
Clearly FSRQs, which generally have very strong emission lines, require 
lower S/N to make a solid ID and thus are more likely to be identified 
than BL Lacs.

In the original X-ray selected sample, we found 39\% of the objects 
to be BL Lacs, along with a significantly lower efficiency for finding 
blazars (86\%).  In the FoM-selected set, we have classifications for 30 
BL Lacs and 143 FSRQs (17.3\% BL Lacs) detected in X-rays.  
The BL Lac 
fraction is much smaller for sources not detected in the RASS: 
we find 38 BL Lacs and 288 FSRQs (11.7\% BL Lacs).  Thus, 
as expected the 
X-ray detection criteria select a significantly larger percentage of 
BL Lac objects.

\section{Conclusions}

We have compiled a survey of 710 objects selected from radio and X-ray 
data to have properties similar to the 3EG blazar counterparts.  
Of these objects, we have contributed 167 new spectroscopic 
identifications along with robust redshifts for the majority.
Our spectroscopy represents an improvement of $\sim$50\% over the 
archival IDs in the northern sky.  Because these sources 
were selected in the same way as the 3EG counterparts but were not 
detected in the 3EG catalog, we believe that many of these sources 
were simply in low states during their \EGRET exposures.  However, we 
cannot rule out the possibility that the differences in the $\gamma$-ray 
and radio beaming may preclude detections of these sources in the GeV 
range.  
We introduce two means of statistically ranking \EGRET $\gamma$-ray 
detections by combining information collected during the 
survey phase of the mission.  Providing further evidence for the 
blazar nature of these sources, we have convincingly detected 
the signal of these FoM-selected candidate blazars in the 
\EGRET data at the $\sim$3.5$\sigma$ level by both of these 
methods.  However, the FoM for individual sources was only 
weakly correlated with the significance of excess $\gamma$-ray flux.

\GLAST will help test whether these 
FoM-selected sources are strong $\gamma$-ray emitters.  In 
addition, we expect that \GLAST will put tighter constraints on 
the typical duty cycle of blazars.  Based on the findings 
in this work, we also believe that a 4EG catalog needs to be created 
to adequately model the blazar population in preparation 
for {\it GLAST}.  Since \GLAST will of course produce an all-sky survey, we 
are extending a CLASS-like VLA survey south of decl. = 0$^\circ$ from 
which we expect to draw a similar number of $\gamma$-ray blazar 
candidates.  Ultimately we seek to identify a nearly uniform all-sky 
set of 2,000-3,000 blazars to match the expected \GLAST detections.

\acknowledgments

DSE and SEH were supported by SLAC under DOE contract DE-AC03-76SF00515, and 
PFM acknowledges support from NASA contract NAS5-00147.

\input{tab1.tex}

\begin{deluxetable}{lrr}
\tablewidth{0pc}
\tablecaption{X-ray Selected Targets}
\tablehead{
\colhead{Name(J2000)}&\colhead{{\it z}} & \colhead{Type$^a$}}
\startdata 
J064501.37+602442.3&0.8319&f \\
J073502.30+711506.4&0.7950&f \\
J074129.61+255733.8&2.7425&f \\
J081108.33+573024.0&0.4180&n \\ 
J085749.78+013531.7&0.2818&b \\ 
J085930.71+621730.4&--&b \\
J090147.03+655150.1&2.0720&f \\
J091925.68+110657.2&0.4232&b \\
\enddata
\tablenotetext{a}{Classification: f = FSRQ, b = BL Lac, n = NLRG, e = Passive 
Elliptical} 
\end{deluxetable}

\begin{deluxetable}{lrrrr}
\tablecolumns{5}
\tablewidth{0pc}
\tablecaption{$\gamma$-ray Probability Distribution}
\tablehead{ & \colhead{\#Prob$<$0.005} & \colhead{$<$0.01} & \colhead{$<$0.05} 
& \colhead{$<$0.1}}
\startdata
Peak	&4	&6	&34	&68\\
Product &0	&8	&41	&79\\
Random 	&2.9	&5.8	&28.8	&57.6
\enddata
\end{deluxetable}

\end{document}

%% file: tab1.tex
\begin{deluxetable}{lrrrrrrccrr}
\tablewidth{0pc}
\tabletypesize{\scriptsize}
\tablecaption{FoM-selected Objects}
\tablehead{
\colhead{Name} & \colhead{FoM} & \colhead{Prod Prob}&\colhead{Peak Prob}&
\colhead{S$_{8.4}$(mJy)} & \colhead{$\alpha$} & \colhead{RASS Cnts}&
\colhead{R2}&\colhead{z} & \colhead{Type$^a$} & \colhead{ID Origin$^b$}
}
\startdata 
J000504.36+542824.9&0.106&--&--&382.5&-0.47&0.000&18.2&0.504&F&2\\
J000520.21+052410.7&0.065&0.489&0.732&228.9&-0.33&0.000&16.2&1.887&F&1\\
J000557.17+382015.1&0.134&0.044&0.084&1077.6&-0.37&0.000&17.1&0.229&F&1\\
J000648.78+242236.4&0.051&0.833&0.915&230.9&-0.14&0.000&18.7&--&?&0\\
J001028.74+204749.7&0.072&0.841&0.836&272.1&-0.31&0.042&18.4&0.600&F&1\\
J001033.98+172418.7&0.074&0.183&0.370&867.5&0.01&0.055&16.9&1.601&F&1\\
J001131.90+704531.6&0.073&--&--&657.1&-0.02&0.000&--&--&?&0\\
J001130.40+005751.8&0.071&0.058&0.345&278.7&-0.28&0.026&20.0&1.492&F&1\\
\enddata
\tablecomments{$\gamma-$ray probabilities are only listed for 
sources that survive our spatial cuts.  Uncertain redshifts are denoted by trailing ':'.}
\tablenotetext{a}{USNO B1.0 R2 magnitude}
\tablenotetext{b}{Classification: F = FSRQ, B = BL Lac, N = NLRG, ? = UNIDED.} 
\tablenotetext{c}{Origin of Spectroscopy: 0 = No info, 1 = Archival, 
2 = Our McDonald Observatory spectroscopy, 3 = Our McDonald Observatory 
spectroscopy of a featureless BL Lac, 4 = Our McDonald Observatory 
spectroscopy in addition to archival data, 5 = Archival featureless BL Lac} 
\end{deluxetable}